\documentclass[aps,pra,twocolumn]{revtex4-1}

\usepackage{graphicx}

\begin{document}

\title{Experimental investigation of the detection mechanism in WSi nanowire superconducting single photon detectors} 



\author{Rosalinda Gaudio$ ^* $}
\affiliation{COBRA Research Institute, Eindhoven University of Technology P.O. Box 513, 5600MB Eindhoven, Netherlands}
\email{r.gaudio@tue.nl\newline $^*$ These authors contributed equally.\newline $^{\dagger}$ Present affiliation: Clarendon Laboratory, Parks Road, Oxford OX1 3PU, United Kingdom.}
\author{Jelmer J. Renema$ ^{*\dagger} $}
\affiliation{Huygens-Kamerlingh Onnes Laboratory, Leiden University, Niels Bohrweg 2, 2333CA Leiden, Netherlands}
\author{Zili Zhou}
\affiliation{COBRA Research Institute, Eindhoven University of Technology P.O. Box 513, 5600MB Eindhoven, Netherlands}
\author{Varun B. Verma}
\author{Adriana E. Lita}
\author{Jeffrey Shainline}
\author{Martin J. Stevens}
\author{Richard P. Mirin}
\author{Sae Woo Nam}
\affiliation{National Institute of Standards and Technology, 325 Broadway, Boulder, Colorado 80305, USA}

\author{Martin P. van Exter}
\author{Michiel J.A. de Dood}
\affiliation{Huygens-Kamerlingh Onnes Laboratory, Leiden University, Niels Bohrweg 2, 2333CA Leiden, Netherlands}
\author{Andrea Fiore}
\affiliation{COBRA Research Institute, Eindhoven University of Technology P.O. Box 513, 5600MB Eindhoven, Netherlands}



\date{\today}

\begin{abstract}
\vspace{2em}
\textbf{We use quantum detector tomography to investigate the detection mechanism in WSi nanowire superconducting single photon detectors (SSPDs).
	To this purpose, we fabricated a 250nm wide and 250nm long WSi nanowire and measured its response to impinging photons with wavelengths ranging from $\lambda = $ 900 nm to $\lambda = $ 1650 nm. Tomographic measurements show that the detector response depends on the total excitation energy only. Moreover, for total absorbed energies $>$ 0.8eV the current-energy relation is linear, similar to what was observed in NbN nanowires, whereas the current-energy relation deviates from linear behaviour for total energies below 0.8eV.}
	
\vspace{4em}
\end{abstract}


\maketitle 

\section{Introduction}
Nanowire superconducting single photon detectors (SSPDs)\cite{goltsman2001} constitute a key technology for the development of quantum communication and computation\cite{natarajan}. Their fast response time, combined with their low dark count rate, low jitter and single- and multi-photon counting capability favours the use of this technology in applications such as quantum key distribution (QKD)\cite{Takesue}, quantum optics\cite{Zinoni}, nanoscale imaging\cite{Bitauld} and interplanetary optical communication\cite{Shaw}. 

Since the first SSPD demonstration, different polycrystalline superconducting films, such as NbN, NbTiN and TaN were employed, and different techniques were developed in order to improve the coupling with incoming light and the photon absorption\cite{natarajan}. Despite technological efforts to improve the device performance, these detectors are still affected by low fabrication yield\cite{kerman,APLGaudio} and the highest system detection efficiency (SDE) reported for $\lambda$ = 1550 nm is not higher than 80$\%$\cite{yamashita_Lowfillingfactor, Rosenberg}. 

Recently, amorphous superconducting films have attracted the interest of the SSPD community\cite{Baek APL, Verma APL, Verma2.5K, WSi}. Although operating at much lower temperatures, SSPDs based on amorphous WSi\cite{WSi} (and also MoSi\cite{Korneeva2013} and MoGe\cite{MoGeVerma}) turned out to be promising for their internal detection efficiency, which saturates close to unity\cite{Baek APL} at currents well below the critical current. Due to their high internal detection efficiency, devices patterned from such amorphous superconducting films exhibit a system detection efficiency higher than 90\%\cite{WSi}. Additionally, devices based on WSi films present more reproducible characteristics, enabling the realization of large arrays \cite{Verma_array}. At present, it is an open question whether these striking differences between NbN and WSi films are related to a fundamentally different nature of the detection process.

For NbN SSPDs, we recently demonstrated \cite{jelmer PRL, polarization} that the detection event is due to a vortex crossing induced by a cloud of quasiparticles which reduces the barrier potential for vortex entry.  The energy dissipated by the vortex crossing the nanowire leads to a transition to the normal state \cite{Semenov2005, Bulaevskii2, Bulaevskii}. Contrary to early models \cite{Semenov2001}, we found that the detection event cannot be described by the local increase of current density over the critical value due to a photo-generated normal core hotspot. 

For WSi, in contrast, little is known about the detection mechanism. Compared to NbN\cite{semenovproperties, bartolfproperties, chockproperties}, a typical \cite{NISTfilms} thin WSi film is characterized \cite{Kondo,Engel} by a higher normal-state electron diffusion coefficient of 0.75 $\mathrm{cm^2 / s}$ vs. $\mathrm{0.5\, cm^2 / s}$, a larger coherence length (9 nm versus 4 nm), a lower superconducting gap (0.5 meV vs. 2 meV) and a lower density of states at the Fermi level ($2 * 10^{22}\, \mathrm{eV^{-1} cm^{-3}}$ vs. $4 * 10^{22}\, \mathrm{eV^{-1} cm^{-3}}$). According to simulations that take into account these properties, these differences are enough to lead to a qualitative change in the detection mechanism. Absorption of a singe photon is expected to result in the formation of a normal hotspot, for a photon of visible, near-infrared, or even mid-infrared wavelength \cite{Engel}. 

Moreover, pump-probe experiments on the two materials produce qualitatively very different results. In WSi\cite{Marsili WSi2}, the lifetime of an excitation created by an absorbed photon is strongly dependent on bias current. In a bias current range from 0.45 to 0.65 $I_b/I_c$, the excitation lifetime changes by an order of magnitude. In contrast, in NbN\cite{Zhouprl, Zhouthesis}, the lifetime is constant over a similar range of bias currents (0.3 - 0.55 $I_b / I_c$). 

For WSi, these experiments were well described by a theory\cite{Kozorezov} in which the recombination of quasiparticles plays a central role. In NbN, in contrast, the evidence points to the fact that quasiparticle multiplication and diffusion set the relevant timescales. Both experimental evidence \cite{jelmerthesis, polarization} and theoretical calculations \cite{Engel} point to a hotspot size of about 20-30 nm in diameter, which leads to an estimated detection time of 2-5 ps, much shorter than the QP recombination time. These results demonstrate that there are substantial differences both in the phenomenology and the theoretical modeling of these materials.

\begin{figure} 
	\includegraphics[width=8 cm]{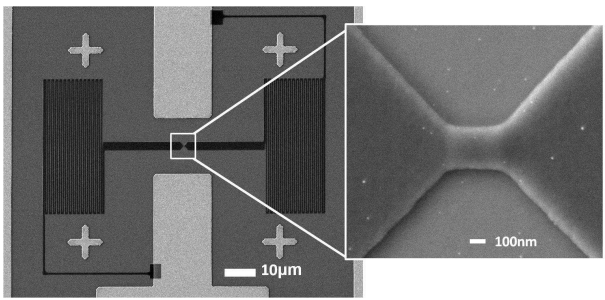}
	\caption{Scanning electron microscope image of the WSi device. The magnification on the right shows the active area of a nanowire detector similar to that involved in the measurements. }
\end{figure}

In this work, we experimentally investigate the nanoscale detection mechanism in WSi nanowire SSPDs. We use quantum detector tomography \cite{Lundeen, RenemaOptExpress} to measure the energy-current relation \cite{JelmerUniversal,renema Sust review} i.e. the amount of bias current ($I_{b}^{th}$) required to produce a detection event with a fixed probability (1\%) as a function of the detected energy ($E_{t}$). This functional dependence is a key signature of the detection mechanism. By the use of multiphoton excitations, the current-energy relation can be measured with sufficient accuracy over a large range of energies to exclude certain classes of models \cite{jelmer PRL}.

We find the same linear scaling between photon energy and bias current at constant detection probability as in NbN SSPDs, which we parameterize as $I_{b}^{th}=I_{o}-\gamma E_{t}$. We observe this scaling in the range $E_{t} = 0.8\mathrm{eV}-2.25\mathrm{eV}$, with a slope $\gamma=1.7\mathrm{\mu A/eV}$, and an extrapolation to zero energy of $I_0/I_c = 0.7$, where $I_c$ is the critical current of the device. As in NbN, we find that the current required to achieve a detection event only depends on the total energy of the photons participating in the detection event. We measure the energy-current relation to sufficient accuracy to exclude the quadratic scaling which is expected from the original normal-core hot spot model \cite{Semenov2001} and from the simulations of Engel \textit{et al}\cite{Engel}. Furthermore, we observe that for energies $E_{t}\leq$0.8eV, experimental data deviate from the linear relation. The strong similarity between our results and those obtained earlier on NbN indicate - surprisingly - that the the differences in material parameters do not substantially alter the phenomenological description of the detection mechanism. 

\section{Fabrication and experimental setup}

The WSi 5nm-thick film is deposited, at room temperature, on a commercial GaAs(001) wafer by co-sputtering W and Si targets. The co-sputtering is performed to obtain a thin film with an estimated composition of $\mathrm{W}_{0.75}\mathrm{Si}_{0.25}$ \cite{Baek APL, Verma APL}. The thin film, characterized by a critical temperature ($T_{c}$) of 3.7K, is then protected against oxidation by a 2nm thick amorphous Si capping layer. The electrical contacts, made of 14nm Ti and 140nm Au layers, are defined by optical lithography and lift off. In the last phase of fabrication, the nanowire is defined via electron-beam lithography (EBL) and patterned via reactive-ion etching in Ar/SF$ _{6} $ plasma. The device design and the EBL exposure were optimized to obtain a nanowire of 250nm width and 250nm length. This width allows us to clearly observe multiphoton detection events, which greatly increases the range of optical energies accessible in the measurements\cite{JelmerUniversal}. We have previously shown\cite{jelmer PRL} for NbN that the current-energy relation is independent of the device geometry: the energy-current relation for a short bridge device is identical to that obtained with a meander. Furthermore, the position dependence of the detection efficiency inferred from such a device carries over to meander devices\cite{polarization}. 

In order to avoid latching\cite{kerman-latching}, the nanowire is defined together with an additional meander (total length 2.8mm and width 300nm) which provides an extra series inductance of approximately 720nH (see figure 1).\newline
The sample is mounted in a VeriCold cryocooler equipped with a final Joule-Thomson stage and is kept at a temperature of 1.6K. At this base temperature the critical current of the WSi nanowire SSPD is $I_{c}$=9.4$\mu$A. During the experiment, the device is biased by a voltage source connected in series with a 10k$ \Omega $ resistor and the voltage drop across it is measured by a multimeter. The DC port of a bias-T (Minicircuits ZNBT-60-1W+) \cite{NISTnote} connects the source and the resistor to the device, while the RF port is connected to a 50$ \Omega $-matched counter. Before reaching the counter, the pulses are amplified by a low-noise amplifier (MITEQ AU-1263) \cite{NISTnote}. To avoid signal reflections, a 4dB attenuator is placed between the bias-T and the amplifier. \newline
The device is illuminated with a Fianium supercontinuum pulsed laser \cite{NISTnote} with repetition rate 20MHz. The laser provides a broadband continuous spectrum from 600nm to 1800nm. For our experiment, the laser beam is linearly polarized perpendicular to the nanowire longitudinal axis and filtered to select the desired wavelength. To keep the polarization axis fixed during the experiment, the polarized light is fed to the device through polarization-maintaining components including optical fibres and a computer controllable digital attenuator. The lensed fibre is mounted inside the VeriCold cryostat and produces a beam spot with nominal diameter of 2.9$\mu$m at 1550nm. 

\begin{figure} 
  \includegraphics[width=8 cm]{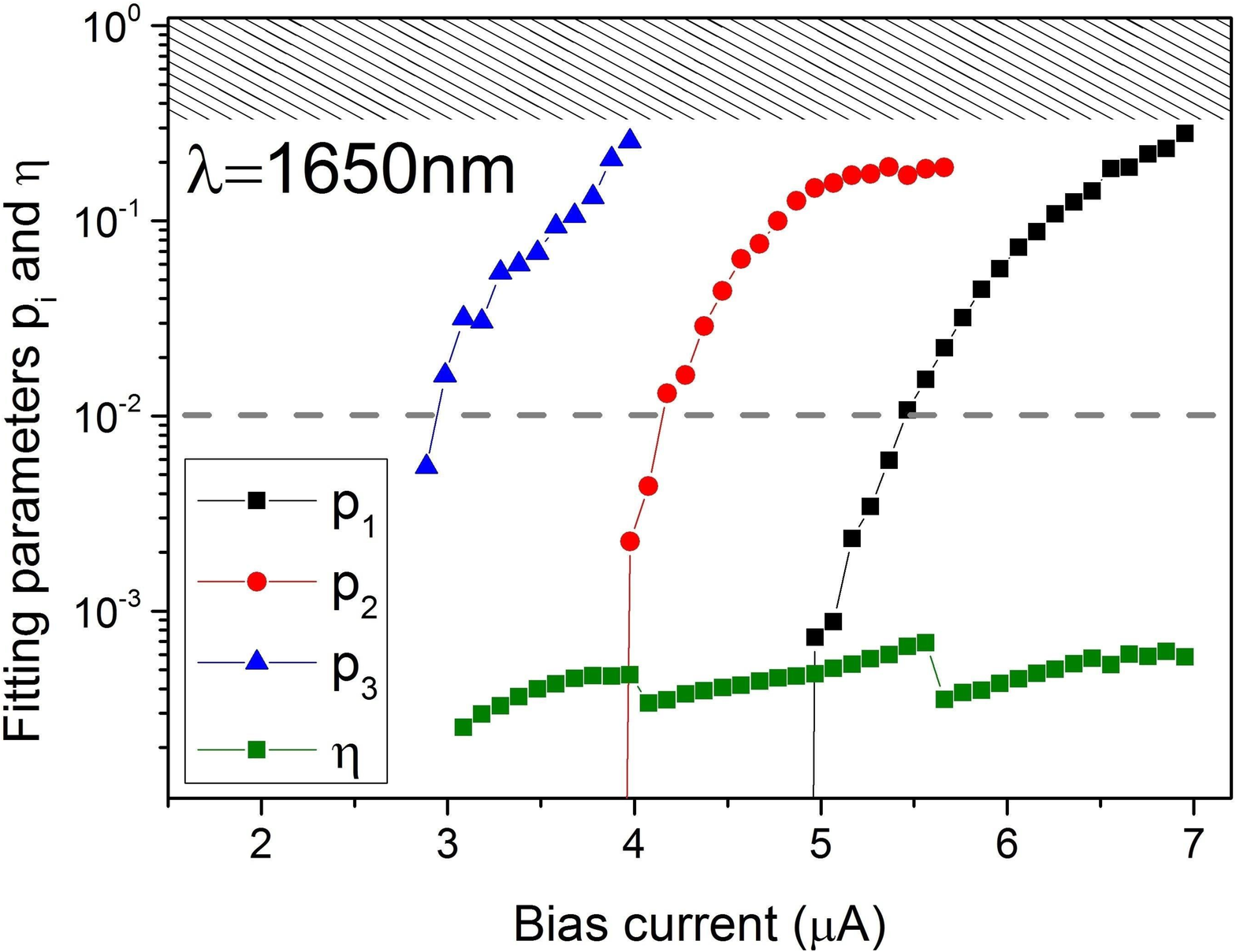}
  \caption{The $p_i$ and $\eta$ displayed as a function of bias current for the case $\lambda$=1650nm. The horizontal dashed line indicates the probability level equal to 0.01. The dashed area indicates the values of $p_n$ which are not accessible with out present measurement.}
\end{figure}
 
\section{Experimental results and discussion}
To measure the multiphoton response of our sample, we make use of quantum detector tomography (QDT)\cite{Lundeen}. The goal of QDT is to measure the probability of a detection event given that $n$ photons are incident on a detector. This is done by recording the detection probability under illumination with coherent states. Since the coherent states form an (overcomplete) basis for the space of quantum states of light, this information is sufficient to infer the response of the device to Fock states (which is the desired quantity) by means of a basis transformation. 

Experimentally, the tomography experiment consists of a series of 11 experimental runs. For each run, a wavelength from the interval 900nm-1650nm is selected with an appropriate bandpass dielectric filter (FWHM 10$\pm$2nm) and is sent through the fiber to illuminate the nanowire. During the experimental run, we record the counts of the detector while varying the light power $P$ and the bias current $I_b$. In order to record the power at each attenuation step, we perform a power calibration after each data acquisition.

We make use of the modified protocol described in ref \cite{RenemaOptExpress}. This protocol uses model selection and an additional sparsity assumption to incorporate all linear loss, regardless of its origin, into a separate effective linear efficiency $\eta$. While this term in principle contains all sources of linear loss, we shall see later that its value is consistent with the absorption probability of the active area of the device. Since the detector is small compared to the wavelengths used in our experiment, this tomography protocol is particularly appropriate for the present detector.  Using this protocol, we obtain the effective linear detection efficiency, $\eta$, as well as the probability that one absorbed photon triggers a detection ($p_1$), two absorbed photons trigger a detection ($p_2$) and so on. The relation between the (classically measurable) mean photon number $N$ and the detection probability $R$ is given by:

\begin{equation}
R(N) = \exp(-\eta N) \sum^{\infty}_{i=0} p_i \frac{(\eta N)^i}{i!}.
\end{equation}

Figure 2 shows a typical data set for an experimental run at one wavelength - in this case 1650 nm. The figure shows the effective linear detection efficiency $\eta$ and the internal detection probabilities $p_1$, $p_2$ and $p_3$ as a function of $I_b$ for the wavelength $\lambda$=1650nm. The probability for $i$ photons to trigger a detection is dependent on $I_b$. For example, for $I_b > 5.5\mu$A the device mostly detects single photons, while for 4$\mu$A $ < $$I_b$$ < $ 5.5$\mu$A it detects predominantly two or more photons ($p_{1}<0.01 $). The data for $I_b > 7\mu$A is not considered since the corresponding pure single-photon regime does not contain any interesting dynamics. 

The observed linear efficiency of $\eta \approx 5 * 10^{-4}$ is consistent with the fraction of photons absorbed into the active area of our detector. The gradual decrease in efficiency at low bias currents could be due to the finite probability of overlap between the excitations along the length of the detector \cite{HSlength, jelmerthesis}. The small jumps in efficiency which occur at $I_b = 4\mu$A and $I_b = 5.5\mu$A are related to the different model (i.e. different number of fitting parameters) used in the different photon-number regimes. This is due to the limited ability of the protocol to resolve values of $p_n \gtrsim 0.3$ due to additional nonlinearities which occur at the high count rates required to resolve such values \cite{Wangtomonoise}. 
 
Once the $p_i$ values for all the wavelengths are obtained, we can find the relation between $I_b^{th}$ and $E_t$ over a wide range of impinging energies. For each wavelength and for each photon regime we record the values of $I_b$ for which the detection probability is equal to 1$\%$ (dashed line in figure 2) and we plot it as a function of the total energy impinging on the detector. The total energy, $E_t$ is $n*E_\varphi$, where $n$ is the photon number and $E_\varphi$ is the energy carried by one photon. This threshold criterion is chosen to be in the range where the imperfections discussed above do not affect our results.

\begin{figure}
  \includegraphics[width=8 cm]{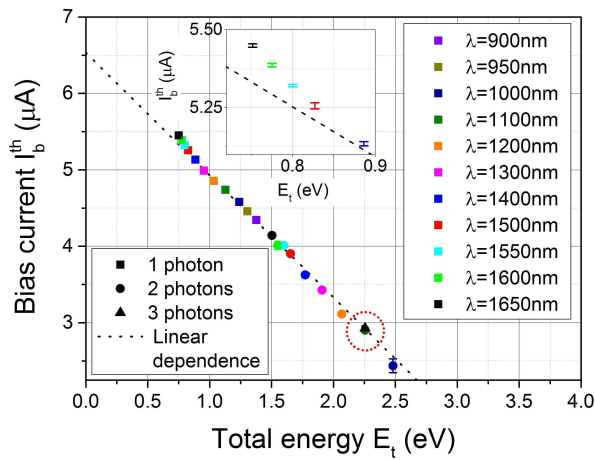}
  \caption{The bias current required to trigger a detection with 0.01 probability is plotted as a function of total energy $E_{t}$ for the 11 wavelengths. The error bars are reported together with data points and range between 2nA and 90nA. The different symbols belong to different detection regimes while each colour corresponds to a wavelength. The red dotted circle highlights the overlap between 2-photon data point for $\lambda$=1100nm and the 3-photon data point for $\lambda$=1650nm. The black dashed line results from a linear fit to the data points with $E_{t} >$0.8eV using the expression $I_{b}^{th}$=$I_{o}$-$\gamma$$E_{t}$. $Inset$: Zoom-in of the upper part of the graph. The data are represented by points to highlight the error bars.}
\end{figure}

Figure 3 shows that the detector responds only to the total excitation energy. Data points corresponding to different wavelengths and numbers of photons lie on the same line, indicating that only the overall excitation energy matters. This is evident from the overlap of two data points highlighted by the red dotted circle in figure 3, corresponding to the detection of three photons with wavelength $\lambda$=1650nm and two photons with $\lambda$=1100nm. We stress that this result is independent of the choice of threshold detection probability, up to a small linear shift, similar to NbN \cite{jelmer PRL}. These results indicate that the detection probability only depends on the total number of photo-created quasi-particles, as was observed in NbN nanowires \cite{jelmer PRL}.
 
The data reported in figure 3 provide the energy-current relation for a WSi nanowire SSPD. For energies corresponding to $E_{t}> 0.8$eV, the data lie on a straight line, which is parameterized by $I_{th} = I_0 - \gamma E_t$, with $\gamma$=1.6$\mu$A/eV and $I_{o}$=6.5$\mu$A=0.7$I_{c}$ (black dashed line in figure 3). A fit of linear behaviour excluding points with $E_{t} < 0.8$ eV gives a significantly better fit ($\chi^2 = 20$) than one which includes low energies ($\chi^2 = 47$). We note that the experimental data are not  well described by the expression $I_{b}^{th}$=$I_{o}$-$\gamma$$\sqrt{E_{t}}$ which characterizes the normal-core hot spot model, regardless of whether we consider the whole data set or only high energies. For NbN, we previously found a linear dependence, with $I_0 = 0.75 I_c$ and $\gamma = 1.6 \mathrm{\mu A / eV}$ for a 220 nm wide detector \cite{jelmer PRL}. In the high energy range, our results are therefore - surprisingly - almost identical to those obtained for NbN. 

However, as shown in the inset, the first three points deviate from this linear trend significantly (as much as 19$\sigma$ for the lowest energy point). We did not observe this deviation in NbN in our previous experiment, which had the same lower energy bound as the present work \cite{jelmer PRL}. It was pointed out previously \cite{lusche width, vodolazov-redBoundary, Kozorezov} for NbN that such a deviation must be expected on physical grounds, since the linear $I_{b}^{th}-E_{t}$ relation cannot hold for $E \approx 0$ if $I_0 < I_c$. If the linear extrapolation would hold to $E = 0$, this would mean that in the absence of impinging energy it would be possible to record a detection event with probability 0.01 if the detector was biased with $I_{b}=I_{o}$, which is not observed in experiments. A preliminary observation of nonlinearity in NbN has recently been reported \cite{vodolazovlatest}.

\section{Discussion}

At present, there are two models which are consistent with our data: the model based on the time-dependent Ginzburg-Landau equation\cite{vodolazov-redBoundary} and the model based on the dynamics of quasiparticle recombination \cite{Kozorezov}. The GL model takes a hotspot of fixed size as its initial condition, and the quasiparticle recombination model takes an area of uniformly suppressed superconductivity as its starting point. It is therefore not surprising that these descriptions work well for WSi, where the hotspot is known \cite{Marsili WSi2} to be larger than in NbN and comparable to the width of the wire. 

The GL-model has the attractive feature that the detection is triggered by the movement of vortices. This ingredient was found to be crucial for explaining the behaviour of NbN devices, because it introduces a dependence on the absorption position, which causes the position-dependent detection efficiency which we demonstrated recently\cite{polarization}. On the other hand, we find no evidence of the low-current detection cutoff which is predicted by the latest version of this model \cite{Vodolazovprivate} More experimental work is needed to determine the detection mechanism in WSi. In particular, it would be interesting to see if WSi has a position dependence, since this would answer the question regarding the role of vortices in the detection mechanism.

\section{Conclusions}

We investigated the detection mechanism in WSi SSPDs. We find that the bias current required to obtain a detection event depends only on the overall excitation energy, not on how that energy is distributed over a number of photons. At high photon energies, we observe a linear dependence between bias current and photon energy required to obtain a detection event. We find that, despite predictions of a normal hotspot in WSi, the square root form of the current-energy relation which is characteristic for some normal-core hotspot models is strongly excluded by our data. We find surprisingly strong similarities between our experimental results on WSi and previous results on NbN.

\section*{acknowledgement}

The authors thank G. Frucci for technical assistance with the experimental setup, J. Francke for assistance during the device wire bonding, and A. Engel, A. Kozorezov, E. Driessen, M. Sidorova, T. Klapwijk and D. Vodolazov for scientific discussions. This work is part of the research programme of the Foundation for Fundamental Research on Matter (FOM), which is financially supported by the Netherlands Organisation for Scientific Research (NWO) and is also supported by NanoNextNL, a micro- and nanotechnology program of the Dutch Ministry of Economic Affairs, Agriculture and Innovation (EL\&I) and 130 partners. J.R. acknowledges support from the NWO Spinoza Prize. 

\section*{References}


\begin{thebibliography}{9}

\bibitem{goltsman2001}
  G. Goltsman, O.Okunev, G. Chulkova, A. Lipatov, A. Semenov, K.Smirnov, B. Voronov, A. Dzardanov, C. Williams, R. Sobolewski,  Appl. Phys. Lett \textbf{79},705, (2001)
  
  
  \bibitem{natarajan}
  C.M. Natarajan, M.G. Tanner and R.H. Hadfield, Supercond. Sci. Technol. $ \textbf{25} $, 063001, (2012).
  
  \bibitem{Takesue}
    H.Takesue, S.W.Nam, Q.Zhang, R.H.Hadfield, T.Honjo, K.Tamaki and Y.Yamamoto, Nature Photonics \textbf{1}, 343 (2007).
    
    \bibitem{Zinoni}
      C. Zinoni, B. Alloing, L.H. Li, F. Marsili, A. Fiore, L. Lunghi, A. Gerardino, Yu B. Vakhtomin, K.V. Smirnov, G.N. Gol’tsman, Appl. Phys. Lett.  \textbf{91}, 031106, (2007).
    
  \bibitem{Bitauld}
   D. Bitauld, F. Marsili, A. Gaggero, F. Mattioli,R. Leoni, S. Jahanmiri Nejad, F. Le´vy, and A. Fiore, Nano Lett.  \textbf{10}, 2977, (2010) 
 
 \bibitem{Shaw}
    M.Shaw, K.Birnbaum, M.Cheng, M.Srinivasan, K.Quirk, J.Kovalik, A.Biswas, A.D. Beyer, F.Marsili, V.Verma, R.P.Mirin, S.W.Nam, J.A.Stern and W.H.Farr, CLEO: Science and Innovations, SM4J. 2 (2014).
    
 
 \bibitem{kerman}
   A.J. Kerman, E.A. Dauler, J.K.W. Yang, K.M. Rosfjord, V. Anant, K.K. Berggren, G. N. Goltsman and B.Voronov, Appl. Phys. Lett. \textbf{90},101110, (2007)
   
 \bibitem{APLGaudio}
 R. Gaudio, K.M.P. op 't Hoog, Z. Zhou, D. Sahin and A. Fiore, Appl. Phys. Lett. \textbf{105}, 222602 (2014).

  \bibitem{yamashita_Lowfillingfactor}
  T. Yamashita, S. Miki, H. Terai and Z. Wang, Optics Express \textbf{21}, 27181 (2013).

\bibitem{Rosenberg}
D. Rosenberg, A. J. Kerman, R. J. Molnar, and E. A. Dauler, Opt. Express \textbf{21}, 1440–1447 (2013).
 
 \bibitem{Baek APL}
   B.Baek, A.E.Lita, V.Verma and S.W.Nam, Appl. Phys. Lett. \textbf{98}, 251105 (2011).
 
 
 \bibitem{Verma APL}
   V.B. Verma, F. Marsili, S. Harrington, A.E. Lita, R.P. Mirin and S.W. Nam, Appl. Phys. Lett.  \textbf{101}, 251114 (2012).
 \bibitem{Verma2.5K}
V.B. Verma, B. Korzh, F. Bussières, R.D. Horansky, A.E. Lita, F. Marsili, M.D. Shaw, H. Zbinden, R.P. Mirin, S.W. Nam, Appl. Phys. Lett. \textbf{105}, 122601,(2014)
    

    \bibitem{WSi}
    F. Marsili, V.B.Verma, J.A.Stern, S.Harrington, A.E.Lita, T.Gerrits, I.Vayshenker, B.Baek, M. D. Shaw, R.P.Mirin and S.W.Nam, Nature Phot., $ \textbf{7} $, 210, (2013)
  

    \bibitem{Korneeva2013}
    Y.P. Korneeva, M.Y. Mikhailov, Y.P. Pershin, N.N. Manova, A.V. Divochiy, Y.B. Vakhtomin, A.A. Korneev, K.V. Smirnov, A.G. Sivakov, A.Y. Devizenko and G.N. Goltsman,  Supercond. Sci. Technol. \textbf{27}, 095012, (2014).
    
       
    \bibitem{MoGeVerma}
    V.Verma, A.Lita, M.R. Vissers, F.Marsili, D.P. Pappas, R.P.Mirin and Sae Woo Nam, CLEO QELS Fundamental Science FM3A 7 (2014)
  
  \bibitem{Verma_array}
  V. B. Verma, R. Horansky, F. Marsili, J. A. Stern, M. D. Shaw, A. E. Lita, R. P. Mirin and S. W. Nam, Appl. Phys. Lett., \textbf{104} 051115, (2014).
    
  
  \bibitem{jelmer PRL}
    J. J. Renema, R. Gaudio, Q. Wang, Z. Zhou, A. Gaggero, F. Mattioli, R. Leoni, D. Sahin, M. J. A. de Dood, A. Fiore and M. P. van Exter, Phys. Rev. Lett. \textbf{112}, 117604 (2014).

\bibitem{polarization}
J. J. Renema, Q. Wang, R. Gaudio, I. Komen, K. op ’t Hoog, D. Sahin, A. Schilling, M. P. van Exter, A. Fiore, A. Engel, and M. J. A. de Dood,  Nano Lett. \textbf{15} 4545 (2015).
    

 \bibitem{Semenov2005}
 A. Semenov, A. Engel, H.W. H\"ubers, K. Il’in, and M. Siegel, Eur. Phys. J. B \textbf{47}, 495 (2005).
 

  \bibitem{Bulaevskii}
   L. N. Bulaevskii, M. J. Graf, C. D. Batista, and V. G. Kogan, Phys. Rev. B \textbf{83}, 144526 (2011).

  \bibitem{Bulaevskii2} 
  L. N. Bulaevskii, M. J. Graf, and V. G. Kogan, Phys. Rev. B \textbf{85}, 014505 (2012). 
    
\bibitem{Semenov2001}
 A.D. Semenov, G.N. Goltsman and A.A. Korneev, Physica C: Superconductivity, \textbf{351}, 349 (2001).

\bibitem{semenovproperties}
A. Semenov, B. Günther, U. Böttger, H.-W. Hübers, H. Bartolf, A. Engel, A. Schilling, K. Ilin, M. Siegel, R. Schneider, D. Gerthsen, and N. A. Gippius, Phys. Rev. B 80, 054510, (2009). 

\bibitem{bartolfproperties}
H. Bartolf, A. Engel, A. Schilling, K. Il’in, M. Siegel, H.-W. Hübers, and A. Semenov, Phys. Rev. B 81, 024502, (2010).

\bibitem{chockproperties}
S. P. Chockalingam, Madhavi Chand, John Jesudasan, Vikram Tripathi, and Pratap Raychaudhuri, Phys. Rev. B 77, 214503, (2008).

 
\bibitem{NISTfilms}
The values quoted here for WSi are those for films produced at NIST.

 \bibitem{Kondo}
 S. Kondo, J. Mater. Res., \textbf{7}, 853 (1992).
  
  \bibitem{Engel}
  A.Engel, J.Lonsky, X.Zhang and A.Schilling, IEEE Trans. Appl. Supercond. \textbf{25}, 220407 (2015).
 
  
 \bibitem{Marsili WSi2} 
F. Marsili, M.J. Stevens, A. Kozorezov, V.B. Verma, C. Lambert, J.A. Stern, R. Horansky, S. Dyer, S. Duff, D.P. Pappas, \textit{et al.}, arXiv preprint arXiv:1506.03129 (2015).
  
    
\bibitem{Zhouprl}
Z. Zhou, G. Frucci, F. Mattioli, A. Gaggero, R. Leoni, S. Jahanmirinejad, T.B. Hoang, A. Fiore, Phys. Rev. Lett., \textbf{110} 133605 , (2013).

\bibitem{Zhouthesis}
Z. Zhou (2014), \textit{Multiphoton detection with superconducting nanowires}, PhD thesis, Eindhoven University of Technology.

\bibitem{Kozorezov}
A.G. Kozorezov, C. Lambert, F. Marsili, M.J. Stevens, V.B. Verma, J.A. Stern, R. Horansky, S. Dyer, S. Duff, D.P. Pappas, A. Lita, M.D. Shaw, R.P. Mirin and S.W. Nam, Phys. Rev. B \textbf{92}, 064504 (2015).

 \bibitem{Lundeen}
  J.S. Lundeen, A. Feito, H. Coldenstrodt-Ronge, K.L. Pregnell, C. Silberhorn, T.C. Ralph, J. Eisert, M.B. Plenio and I.A. Walmsley, Nature Physics \textbf{5}, 27 (2008).

 \bibitem{RenemaOptExpress}
 J. J. Renema, G. Frucci, Z. Zhou, F. Mattioli, A. Gaggero, R. Leoni, M. J. A. d. Dood, A. Fiore, and M. P. van Exter, Opt. Express \textbf{20}, 2806 (2012).

\bibitem{JelmerUniversal}
    J.J. Renema, G. Frucci, Z. Zhou, F. Mattioli, A. Gaggero, R. Leoni, M.J.A. de Dood, A. Fiore and M.P. van Exter, Phys. Rev. B \textbf{87}, 174526 (2013).
      
 \bibitem{renema Sust review}
 A. Engel, J.J. Renema, K. Il’in and A. Semenov, Superconductor Science and Technology, \textbf{28} 114003 (2015).
 
 \bibitem{kerman-latching}
   A.J. Kerman, E.A. Dauler, W.E. Keicher, J.K.W. Yang, K.K. Berggren, G. N. Goltsman and B.Voronov, Appl. Phys. Lett. \textbf{88},111116, (2006)

\bibitem{NISTnote}
The use of trade names is intended to allow the measurements to be appropriately interpreted, and does not imply endorsement by the US government, nor does it imply these are necessarily the best available for the purpose used here.
       
\bibitem{HSlength}
        J. J. Renema, R. Gaudio, Q. Wang, Z. Zhou, A. Gaggero, F. Mattioli, R. Leoni, D. Sahin, M. P. van Exter, A. Fiore and M. J. A de Dood, $unpublished$.


\bibitem{Wangtomonoise}
 Q. Wang, J. J. Renema, A. Gaggero, F. Mattioli, R. Leoni, M. P. van Exter, M. J. A. de Dood, J. Appl. Phys. \textbf{118}, 134501 (2015)


\bibitem{lusche width}
  R. Lusche, A. Semenov, K. Ilin, M. Siegel, Y. Korneeva, A. Trifonov, A. Korneev, G. Goltsman, D. Vodolazov, H.-W. H\"ubers, Journal of Applied Physics, \textbf{116}, 043906 (2014).
  
\bibitem{jelmerthesis}
 J.J. Renema (2015), \textit{The physics of nanowire superconducting single-photon detectors}, PhD thesis, Leiden University. 
 

\bibitem{vodolazov-redBoundary}
  D. Yu. Vodolazov, Phys. Rev. B., \textbf{90}, 054515 (2014).

\bibitem{vodolazovlatest}
D. Yu. Vodolazov, Yu. P. Korneeva, A. V. Semenov, A. A. Korneev, and G. N. Goltsman, Phys. Rev. B 92, 104503, (2015).

\bibitem{Vodolazovprivate}
 D. Yu. Vodolazov, private communication.
 

    
 \end{thebibliography}
\end{document}